\documentclass[aps,prd,10pt,twocolumn,floatfix,notitlepage,
superscriptaddress,nofootinbib,longbibliography]{revtex4-2}
\usepackage{amsmath,amssymb,booktabs,array,graphicx,xcolor,hyperref,microtype}
\hypersetup{hidelinks,pdftitle={Jet-by-jet fluctuations of the energy-energy correlator: hadronization, decays, and medium effects},pdfauthor={J. Y. Zhang}}
\newcommand{\pt}{p_{\mathrm T}}
\newcommand{\Tr}{\operatorname{Tr}}
\newcommand{\PythiaTraceResult}{Hadronization raises the
summed density variance by roughly a factor of three, and
decays then reduce it by about half. In 50--60 GeV, the
ratios are \(T_{H_0}/T_P=2.84\pm0.49\) and
\(T_H/T_P=1.46\pm0.20\). The dominant hadronization
increase comes from the small-angle diagonal variances;
the decrease at larger angles is much smaller.}
\newcommand{\PythiaRankResult}{In 50--60 GeV, these large
amplitude changes coexist with similarly broad variance
spectra at the three stages. Their effective ranks are about
18--19, with differences comparable to the
statistical uncertainties.}

\begin{document}
\raggedbottom
% Compact only the figure-to-text gap after adding bibliography citations.
\setlength{\dbltextfloatsep}{10pt plus 2pt minus 2pt}
\title{Jet-by-jet fluctuations of the energy-energy correlator:\texorpdfstring{\\}{ }
hadronization, decays, and medium effects}
\author{J. Y. Zhang}
\email{jingyuzhang21@m.fudan.edu.cn}
\affiliation{Key Laboratory of Nuclear Physics and Ion-beam Application
(MOE), Institute of Modern Physics, Fudan University, Shanghai 200433, China}

\begin{abstract}
Energy correlators probe how the strong interaction shapes the energy
distribution within jets. Here, we study the jet-by-jet fluctuations of
the energy-energy correlator (EEC). Each jet's EEC is described by a
vector of densities in angular shells. Using an ensemble of these
vectors, we construct the covariance matrix \(C\) and Pearson
correlation matrix \(R\) to quantify fluctuation amplitudes and
correlations between angular shells. Reconstructed jets in CMS Open
Data at 7 TeV show positive correlations between nearby outer-angle
bins and negative correlations between separated angular regions.
We compare PYTHIA jets at the parton, undecayed-hadron, and final-particle
stages to identify correlations already present in the shower and study
how hadronization and decays reshape the angular fluctuations. PYTHIA is then coupled to linear Boltzmann
transport (LBT) to study medium effects on jet-by-jet EEC
fluctuations and their angular correlations.
These results open a route to probing the stochastic evolution of
QCD jets through jet-by-jet EEC fluctuations.
\end{abstract}
\maketitle

\section{Introduction}

Jets provide a way to study quantum chromodynamics (QCD), from
parton radiation to hadron formation. Energy correlators describe
jet substructure through momentum-weighted particle pairs and
higher multiplets \cite{Basham:1978bw,Larkoski:2013eya,Chen:2020vvp}.
The energy-energy correlator (EEC) measures how these particle-pair
contributions are distributed over angular separation \cite{Moult:2025nhu}.
Calculations of two- and higher-point correlators
\cite{Dixon:2019uzg,Chen:2019bpb,Chen:2022jhb,Yang:2022tgm}
and measurements in jets \cite{CMS:2024mlf,ALICE:2024dfl,STAR:2025jut}
have developed this approach. CMS Open Data also allow the EEC to
be studied in jets reconstructed from recorded proton-proton collisions
\cite{Komiske:2022enw,Komiske:2019fks}.

Event-by-event distributions of integrated energy correlators have
long been used for jet discrimination \cite{Larkoski:2013eya}.
Related studies examine multiplicity-dependent EECs \cite{Duan:2026icj},
particle correlations within jets \cite{Zhao:2025ogc}, and correlations
among energy observables \cite{Bal:2026triality}.
Here we retain the angular EEC profile of each jet and ask which
angular regions fluctuate together and which vary in opposite directions.
Its covariance describes these variations across the jet ensemble,
rather than the statistical uncertainty of the mean EEC.
For distinct angular bins, the covariance combines shared-constituent
three-particle terms and four-distinct-particle terms, with the
product of the mean EEC densities subtracted. This connects the
observable to multiparticle energy correlations
\cite{Chen:2022swd,Komiske:2017aww,Bal:2026triality,Gonzalez:2026multipoint},
projected onto the two measured pair angles.

We first use CMS Open Data to study the distribution of EEC
profiles across jets, their angular covariance, and the variance
spectrum. We then compare jets reconstructed at three stages of
Pythia evolution to ask how hadronization and decays change the
covariance. These processes
share momentum among new particles and change many pairs at once.
They complement the effects studied in mean correlators through
power corrections, track functions, and fragmentation
\cite{Schindler:2023cww,Lee:2024esz,Li:2021zcf,Jaarsma:2022kdd,Liu:2024lxy,Chen:2024nyc,Jaarsma:2023ell}.
A separate same-history comparison follows matched jets before and
after decays and decomposes the change in their summed shell variance.
Medium effects have also been studied with two-point
\cite{CMS:2025ydi,Andres:2024ksi,Singh:2024vwb,Andres:2024hdd} and
higher-point energy correlators \cite{Bossi:2024qho,Barata:2025fzd}.
A partonic medium changes the momenta and color connections that
feed this evolution. Matched vacuum and medium histories from Pythia
and linear Boltzmann transport (LBT) track changes in both fluctuation
strength and angular correlations through hadronization and decays.

\begin{figure*}[t]
 \centering\includegraphics[width=.94\textwidth]{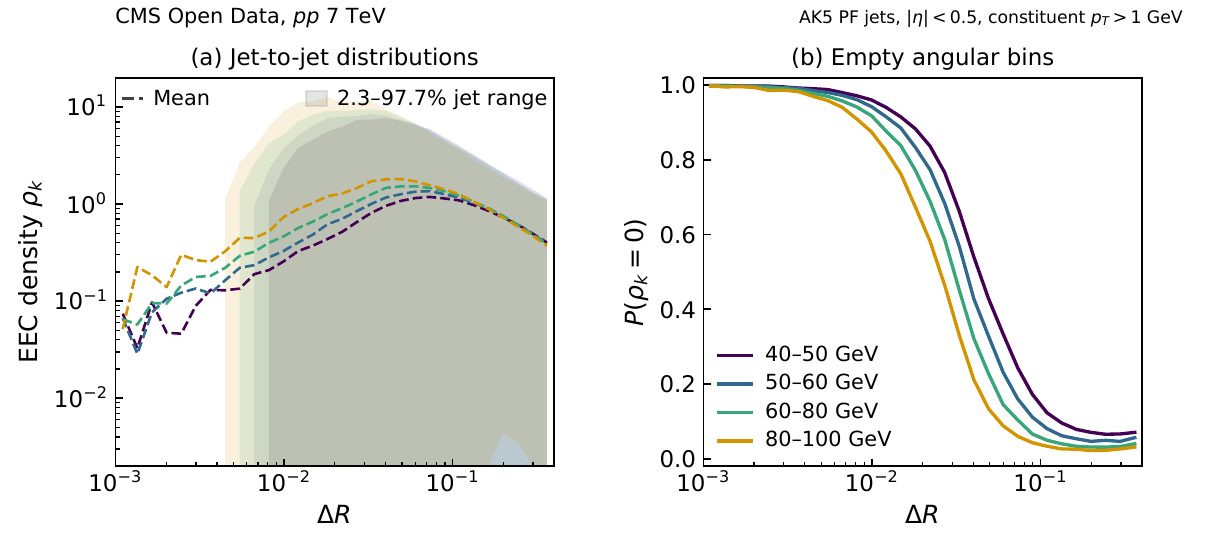}
 \caption{EEC-density distributions in reconstructed CMS Open Data
 jets at 7 TeV. (a) Mean (dashed) and 2.3--97.7\% jet quantiles
 (shaded), including empty bins; bands reaching zero extend to
 the lower plotting limit. (b) Empty-bin probability with
 16th--84th percentile bootstrap intervals. Colors denote
 corrected jet-momentum intervals.}
 \label{fig:cms-distribution}
\end{figure*}

\section{Observable and analysis samples}
\label{sec:observable}

For each selected jet, we divide the pair-angle window into
nonoverlapping bins \(B_k\), with linear widths
\(w_k=\Delta R_k^{\rm high}-\Delta R_k^{\rm low}\). The bin-averaged
EEC density is
\begin{equation}
 \rho_{J,k}=\frac{1}{w_k}\sum_{i<j\in J}z_i z_j
 \mathbf{1}(\Delta R_{ij}\in B_k),\qquad
 z_i=\frac{p_{T,i}}{p_{T,J}^{\rm pre}}.
 \label{eq:rho}
\end{equation}
Here \(\Delta R_{ij}=\sqrt{(\eta_i-\eta_j)^2+(\Delta\phi_{ij})^2}\)
is the constituent-pair separation, with the azimuthal difference
wrapped into \([-\pi,\pi)\). The sum counts unordered pairs once and omits
self-correlations. Its ensemble mean is the differential EEC in
this convention. The density is defined per unit \(\Delta R\),
also for logarithmic bins. The integrated bin contribution is
\(M_{J,k}=w_k\rho_{J,k}\).

The covariance and Pearson correlation are
\begin{align}
 \mu_k&=\langle\rho_k\rangle,\nonumber\\
 C_{k\ell}&=\langle(\rho_k-\mu_k)(\rho_\ell-\mu_\ell)\rangle,
 \label{eq:covariance}\\
 R_{k\ell}&=\frac{C_{k\ell}}{\sqrt{C_{kk}C_{\ell\ell}}}.
 \label{eq:pearson}
\end{align}
The brackets denote an average over selected jets. For CMS data,
\(\langle f\rangle=\sum_Ju_Jf_J/\sum_Ju_J\), where \(u_J\) is
the trigger-prescale weight of the collision. Generator jets have
equal weights. Each sample uses its own mean profile.

Positive covariance means that the EEC densities in two shells
tend to rise or fall together across jets; negative covariance
means that they tend to vary in opposite directions. The Pearson
matrix expresses this dependence relative to the standard deviation
in each shell. We evaluate \(R_{k\ell}\) for shells with nonzero variance.

We analyze the CMS 2011A Jet AOD dataset at \(\sqrt{s}=7\) TeV
\cite{CMS:OpenData2011Jet}. Jets are reconstructed from particle-flow
(PF) candidates \cite{CMS:PF2017} with the anti-\(k_T\) algorithm
\cite{Cacciari:2008gp} and radius \(R_{\rm jet}=0.5\).
We retain the two jets with highest uncorrected transverse momentum
in each event, then require medium jet identification,
\(|\eta_J|<0.5\), and corrected jet momenta in 40--50, 50--60,
60--80, or 80--100 GeV. The selected sample contains about 55\,000 jets.
Events pass certified luminosity sections, a primary-vertex selection,
and a version of \texttt{HLT\_DiJetAve30U}. The product of the
applicable L1 and HLT prescales weights each selected jet.
These are reconstructed, trigger-selected distributions; no
trigger-efficiency correction or detector unfolding is applied.
Appendix~\ref{app:samples} gives the sample definitions.

\begin{figure*}[t]
 \centering\includegraphics[width=.86\textwidth]{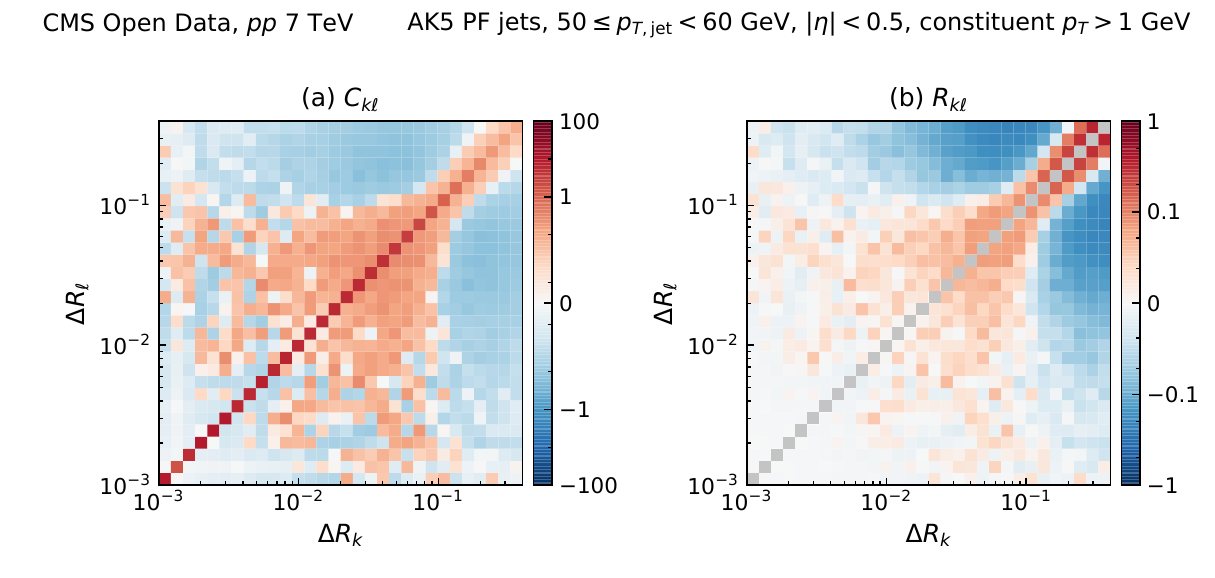}
 \caption{EEC-density covariance \(C\) (a) and Pearson correlation
 \(R\) (b) for 50--60 GeV jets in CMS Open Data. Color scales are
 symmetric logarithmic for \(C\) and inverse hyperbolic sine for
 \(R\). The unit diagonal of \(R\) is gray; undefined coefficients
 are white.}
 \label{fig:cms-matrix}
\end{figure*}

The EEC uses all PF constituents with \(p_T>1\) GeV after jet
reconstruction. Its denominator is the full uncorrected jet
momentum before that cut; the jet energy correction is used only
for jet selection. We measure particle-pair separations in
\(0.001\leq\Delta R<0.4\), using 30 logarithmic bins.
Pairs outside this window are omitted from the EEC, while the
full jet momentum remains in its normalization.
The constituent threshold selects which particles enter the pair sum.
A collinear splitting can change the retained contribution if its
daughters fall below the cut.

\begin{figure*}[t]
 \centering\includegraphics[width=.98\textwidth]{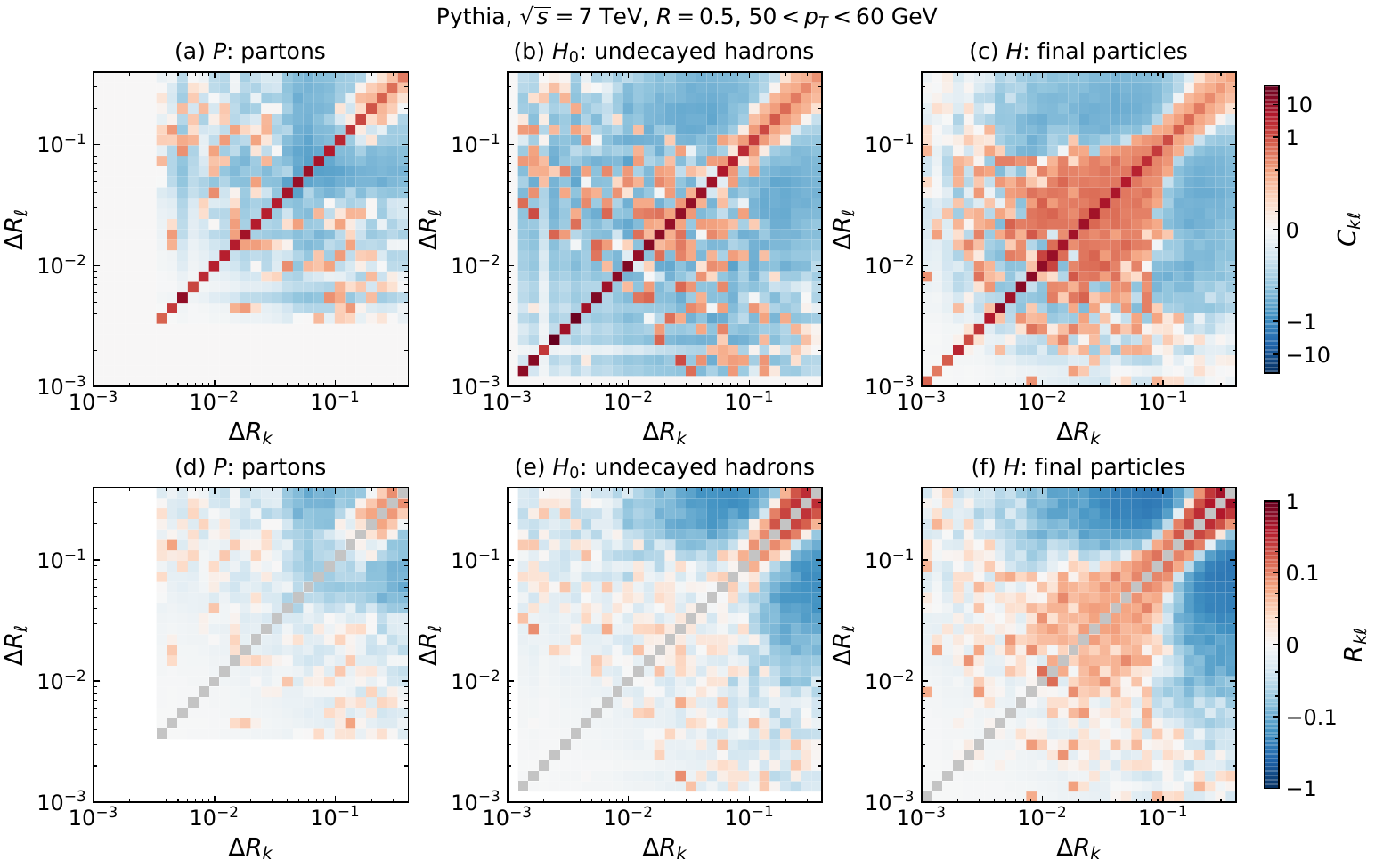}
 \caption{Pythia EEC-density covariance \(C\) (top) and Pearson
 correlation \(R\) (bottom) for 50--60 GeV jets at 7 TeV and
 \(R_{\rm jet}=0.5\). Columns show partons \(P\), undecayed
 hadrons \(H_0\), and final particles \(H\). Color scales are
 common within each row; axes and diagonal conventions follow
 Fig.~\ref{fig:cms-matrix}.}
 \label{fig:pythia-matrix}
\end{figure*}

\begin{figure*}[t]
 \centering\includegraphics[width=.80\textwidth]{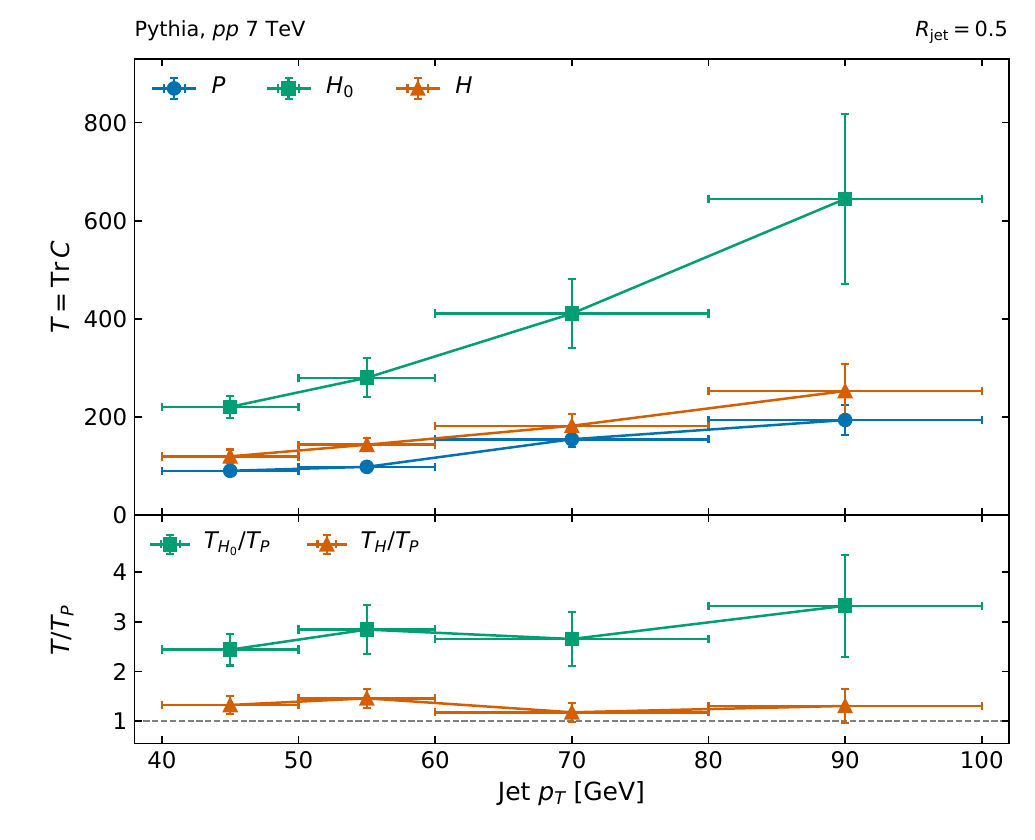}
 \caption{Pythia EEC-density variance trace \(T=\mathrm{Tr}\,C\)
 for \(P,H_0,H\) (top) and ratios \(T_{H_0}/T_P\),
 \(T_H/T_P\) (bottom). The dashed line marks unity. Horizontal
 bars span the jet-momentum intervals; vertical bars give
 bootstrap standard deviations.}
 \label{fig:pythia-trace}
\end{figure*}

The Pythia comparison uses 7 TeV collisions, the same jet radius,
selection from the two highest-\(p_T\) jets in each event, acceptance,
momentum bins, constituent cut,
and EEC grid. Jet selection there uses the generated jet momentum.
The stages are final partons \(P\), hadrons with subsequent decays
disabled \(H_0\), and final visible particles with standard decays
\(H\). Figures~\ref{fig:pythia-matrix} and~\ref{fig:pythia-trace}
compare independently generated jet ensembles at these three stages.
Jets are reconstructed and selected separately at each stage using
the same criteria. Their differences include changes in clustering
and sample selection.
Appendix~\ref{app:samples} specifies the generator settings.

To follow the decay response of matched jets, we use a separate
same-history Pythia sample and save each event before and after the standard
decay chain. Jets are reconstructed separately at the two
stages. In each event, we first retain the two highest-\(p_T\)
\(H_0\) jets, then select reference jets with
\(50\leq\pt<60\) GeV and \(|\eta_J|<0.5\). They are matched to
mutually nearest \(H\) jets in \((\eta,\phi)\) within
\(\Delta R_{\rm jet}<0.25\), without reapplying these selections to
the partner. Both EEC profiles use the definition in Eq.~\eqref{eq:rho}.

Momentum sharing and changes in particle-pair separations can alter
both the size and angular shape of a jet's EEC profile. We use the
covariance spectrum to study whether the jet-by-jet variation follows
one dominant angular pattern or is spread over many patterns. The eigenvectors of
\(C\) specify orthogonal angular patterns; each eigenvalue is the
variance of the EEC profiles projected onto its unit eigenvector.
Ordering the eigenvalues from largest to smallest gives the variance
spectrum. We define
\begin{equation}
 \begin{gathered}
 T=\Tr C=\sum_a\lambda_a,\qquad p_a=\frac{\lambda_a}{T},\\
 r_{\rm eff}=\exp\!\left(-\sum_a p_a\ln p_a\right),
 \end{gathered}
 \label{eq:spectrum}
\end{equation}
with \(0\ln0=0\) and the last two quantities defined for \(T>0\).
On this fixed grid, \(T\) sums the shell-by-shell density variances,
and \(p_a\) is the fraction carried by mode \(a\). The effective rank
\cite{Roy:2007effective}
measures how broadly this variance is shared among modes: it is one
when a single mode carries all the variance and increases when more
modes carry comparable shares.
A common rescaling of all EEC profiles changes \(T\), but not
\(p_a\) or \(r_{\rm eff}\). These quantities distinguish changes in
fluctuation strength from changes in its distribution among angular modes.
The variance of the angle-integrated EEC is instead
\(\operatorname{Var}(\sum_kw_k\rho_k)=w^TCw\).

We estimate statistical uncertainties by bootstrap resampling
of collisions, keeping their jets together and recomputing the
observables. Independent samples are resampled separately;
matched stages and vacuum--medium partners are resampled together.
The asymmetric errors on the matched-decay trace ratio denote
a 95\% bootstrap interval.
The medium comparisons use 95\% percentile intervals.

\begin{figure*}[t]
 \centering\includegraphics[width=.94\textwidth]{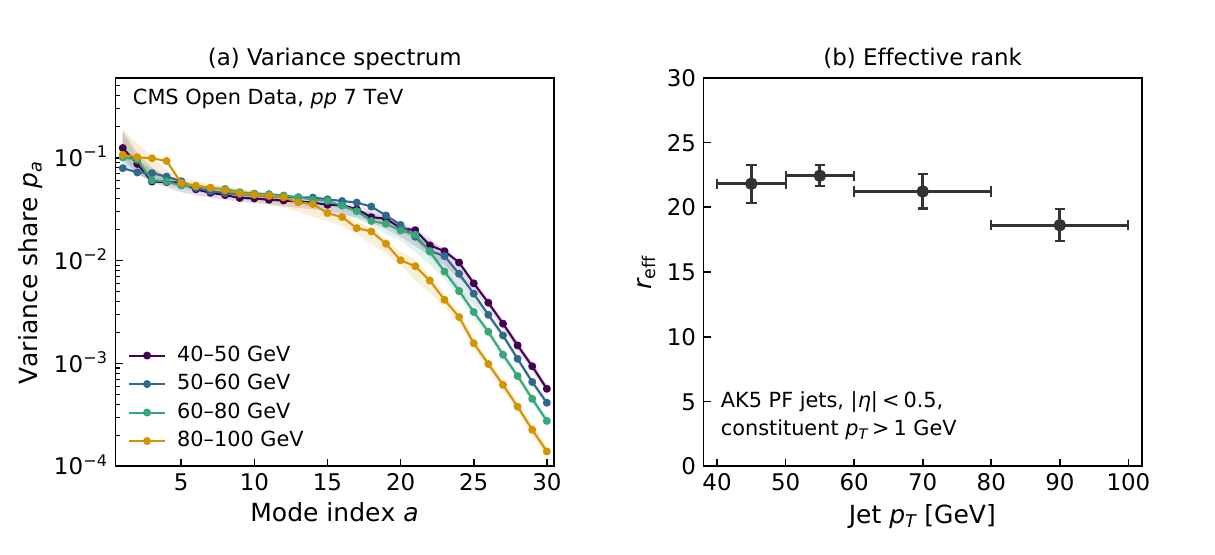}
 \caption{CMS EEC-density variance spectrum. (a) Variance shares
 \(p_a\), ordered by decreasing eigenvalue, with 16th--84th
 percentile bootstrap bands; zero eigenvalues are omitted.
 (b) Entropy effective rank versus corrected jet momentum.
 Horizontal bars span the momentum intervals; vertical bars give
 bootstrap standard deviations.}
 \label{fig:cms-spectrum}
\end{figure*}

\section{Jet-by-jet fluctuations in CMS Open Data}
\label{sec:data}

We first examine how the EEC density in each angular shell varies
from jet to jet. Figure~\ref{fig:cms-distribution} shows these
distributions through their mean, quantile bands, and empty-bin
probability. At small separations, most jets have no selected pair
in a shell, yet the mean EEC is nonzero.
The distribution has a finite probability at zero and a tail of
positive contributions, rather than Gaussian fluctuations around
the mean. The broad quantile bands show the size of the jet-to-jet
variation. The smooth mean profile therefore combines many
empty shells with occasional large contributions. Higher-momentum
jets populate the small-angle shells more often, and their mean-EEC
maximum moves to smaller separations.

The sparse, non-Gaussian distributions show that a smooth mean EEC
can conceal large differences between individual jets. Retaining
each jet's profile lets us study these differences and their
correlations across angles.
The empty-bin probability depends on the angular resolution and
constituent cut: narrower shells capture fewer pairs.
We next ask which angular contributions tend to
appear together and which compete across the jet ensemble.

The mean EEC provides a physical reference for these angles.
At larger separations, its shape reflects parton branching; at
small separations, hadronic degrees of freedom dominate. Between
them, the turnover marks the region where hadronization becomes
important \cite{Komiske:2022enw,ALICE:2024dfl}. Relative to the
broad maximum in Fig.~\ref{fig:cms-distribution}, separations above
about \(0.1\) lie on the larger-angle, shower-dominated side.
The interval \(0.01\)--\(0.1\) spans the turnover and extends
onto its small-angle, hadronic side. The moving maximum provides
the reference as the jet momentum changes.

We locate these correlations across pair angles using the \(C\) and
\(R\) matrices in Fig.~\ref{fig:cms-matrix}. Negative correlations
connect the region around and below the mean-EEC turnover to the
larger-angle region shaped by the parton shower. These broad
off-diagonal regions are clearest in the Pearson matrix. Across
jets, an enhanced EEC contribution on one side tends to accompany
a reduced contribution on the other.
Nearby shells above
\(\Delta R\simeq0.1\) instead fluctuate together. A weaker
positive near-diagonal band is also present between approximately
\(0.01\) and \(0.1\), around and below the mean-EEC maximum.
The shell variances differ greatly, so we use the Pearson matrix
to study the local correlations between their fluctuations.

These patterns distinguish competition between separated angular
regions from common changes across nearby ones. Momentum sharing
provides a simple picture of the negative correlation: redistributing
momentum among branches or changing their opening angles can
increase the EEC at one separation while reducing it at another.
A branch can also contribute pairs to several nearby shells, which
then respond together to changes in that branch. The smaller-angle
positive band shows that joint fluctuations also extend into the
region shaped by hadron formation and decays. The separated-angle
negative region and the nearby outer-angle positive band persist
across the jet-momentum and constituent-threshold checks in
Appendix~\ref{app:checks}. The Pythia comparison below examines
which structures are already present in the shower and how they
change when hadrons form and decay.

We next ask whether these angular fluctuations are driven by one
or a few common patterns. The eigenvalue spectrum of \(C\) in
Fig.~\ref{fig:cms-spectrum} shows that the variance is instead
shared among many angular modes. The leading mode carries only about
8\% for 50--60 GeV jets, and the effective rank stays near 20 across
the measured momentum intervals. A fixed EEC shape with only an
overall amplitude fluctuation would give rank one. The broad spectrum
instead reflects changes in where pair contributions lie across
angles, as well as in the overall EEC size.

\section{Hadronization and decay effects in Pythia}
\label{sec:stages}

To see which correlations are already present in the shower and
which develop later, we compare three Pythia stages: final
partons \(P\), hadrons after hadronization with decays disabled
\(H_0\), and final visible particles after decays \(H\).
Figure~\ref{fig:pythia-matrix} shows their covariance and Pearson
matrices for the same jet-momentum interval and angular grid as
the CMS example. The negative correlations between the smaller-
and larger-angle regions are already present in the parton
sample. The negative regions become more extensive after
hadronization and remain after decays. The shower therefore
already contains a competition between angular contributions that
is also seen in the final-particle ensemble. The outer neighboring
shells show a positive tendency at parton level and a clearer band
after hadronization. The parton shower thus provides a common
starting point for both structures.

Momentum sharing gives a simple picture of the negative region.
Consider two narrow branches separated by a larger angle, with
identical internal structure and momentum shares \(x\) and \(1-x\).
Keep their total scalar transverse momentum, EEC normalization,
and internal momentum fractions and angles fixed.
Within-branch pairs populate smaller separations; cross-branch
pairs populate larger ones. Since each pair contributes a product
of two particle momenta,
\[
 M_{\rm within}\propto x^2+(1-x)^2,
 \qquad M_{\rm between}\propto x(1-x).
\]
More balanced sharing increases \(x(1-x)\).
For internal pairs, the decrease in the stronger branch's contribution
exceeds the increase in the weaker branch's, since
\(x^2+(1-x)^2=1-2x(1-x)\).
Across jets, these smaller- and larger-angle contributions then
fluctuate oppositely, giving negative correlations.

The same momentum budget links successive branchings. If the
branch carrying \(1-x\) splits with fractions \(y\) and \(1-y\),
the pair it produces has a contribution proportional to
\((1-x)^2y(1-y)\). For fixed \(y\) and \(0<x<1/2\), increasing
\(x\) strengthens the cross-branch contribution but weakens this
internal contribution. The later splitting carries a smaller EEC
contribution because its parent has less momentum.
Changes in branch angles provide another source of negative
correlation. Moving a group of pairs to a larger separation leaves
a deficit at its previous location and an excess at its new one.

Positive correlations can arise within the same picture. Cross-branch
pairs occupy a range of separations around the branch opening angle.
With fixed geometry, consider two nearby shells populated only by
these pairs. Their densities are \(\rho_k=g_kq\) and
\(\rho_\ell=g_\ell q\), where
\(q=x(1-x)\) and the positive constants \(g_k,g_\ell\) include
the angular profiles and bin widths. These contributions have
\[
 \operatorname{Cov}(\rho_k,\rho_\ell)
 =g_kg_\ell\operatorname{Var}(q)>0
\]
when \(q\) fluctuates. Thus neighboring shells can rise and fall
together while competing with a more distant angular region.
Pairs sharing a constituent provide another common source:
\(z_i z_j\) and \(z_i z_m\) respond together to a change in
\(z_i\) when the partner momenta and pair angles are fixed.
These mechanisms give a physical picture of the local positive
band on the larger-angle, parton-sensitive side of the EEC.

Hadronization changes how these partonic patterns are carried by
particle pairs. In the string model, the color-connected parton
system supplies a common momentum budget for the produced hadrons.
String fragmentation shares this momentum and gives the hadrons
momenta transverse to the string
\cite{Bierlich:2022pfr,Skands:2014pea}. It creates new pairs and
changes the separations of pairs that span different branches.
The hadronic EEC can therefore retain the earlier angular structure
while changing its strength and width.

The region around and below the mean-EEC turnover responds
differently. After decays, a broad positive band forms between
approximately \(0.01\) and \(0.1\), with both the neighboring-shell
covariance and Pearson coefficients increasing. This band spans the hadronic side and the turnover
of the mean EEC. The stage comparison thus
connects its local joint fluctuations to the formation and decay
of hadrons, in addition to the correlations inherited from the shower.

A decay links several angular contributions through one parent \(p\).
It adds daughter--daughter pairs and replaces each parent--external
pair with several daughter--external pairs. In the collinear limit,
with fixed normalization and all daughters retained,
\(\sum_a z_a z_j=z_pz_j\), where \(j\) is an external constituent.
Finite decay angles spread the parent's contribution across nearby
shells. Decays thus add pairs to the hadronic angular region.
Their shared origin can make several shell contributions rise
together across jets, providing a simple picture of the positive
correlations near and below the turnover. The constituent cut
selects which pairs enter these shells.

The angular and stage comparisons give complementary views of
the same evolution. Large-angle fluctuations already reflect the
partonic branching pattern. Hadron formation changes how that
pattern is carried by particle pairs, and decays add a pronounced
local correlation band on the hadronic and turnover side of the
EEC. Because daughter--external pairs span several separations,
decays also modify correlations at larger angles. The final matrix
therefore contains both structures inherited from the parton shower
and structures reshaped by hadronization and decays.
At larger separations, decays generally reduce the covariance between
neighboring shells while increasing their Pearson correlation.

\begin{figure*}[t]
 \centering\includegraphics[width=.98\textwidth]{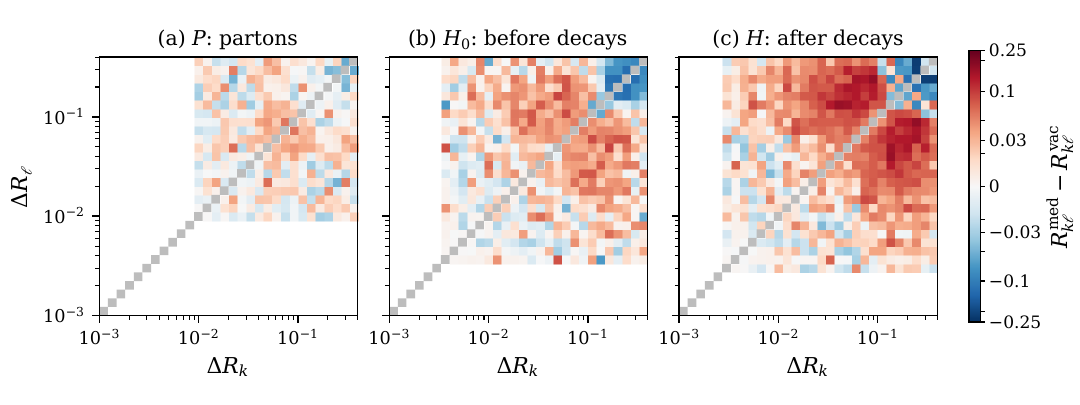}
 \caption{Medium-induced Pearson-correlation change,
 \(\Delta R_{k\ell}=R^{\rm med}_{k\ell}-R^{\rm vac}_{k\ell}\),
 for matched jets at \(P\) (a), \(H_0\) (b), and \(H\) (c).
 Panels share an inverse-hyperbolic-sine color scale. Red denotes
 an increase and blue a decrease; the zero diagonal is gray.
 White cells are undefined or have fewer than 20 nonzero source
 events in either shell in either branch.}
 \label{fig:medium-correlation}
\end{figure*}

We next ask how the total size of the fluctuations changes between
stages. The trace \(T\) sums the individual-bin density variances;
Fig.~\ref{fig:pythia-trace} shows its value and the ratios of both
hadron stages to the parton baseline. \PythiaTraceResult\
\PythiaRankResult\

The same-history sample resolves this decay response by following
matched jets. With \(d_k=\rho_{H,k}-\rho_{H_0,k}\), its trace change is
\begin{equation}
 \begin{aligned}
 T_H-T_{H_0}&=\sum_k\operatorname{Var}(d_k)\\
 &\quad+2\sum_k\operatorname{Cov}(\rho_{H_0,k},d_k).
 \end{aligned}
 \label{eq:paired-decay-trace}
\end{equation}
The matched sample gives \(T_H/T_{H_0}=0.49^{+0.18}_{-0.13}\):
decays reduce the summed shell variance by about half.
The negative cross term in Eq.~\eqref{eq:paired-decay-trace} outweighs
the variance of the decay response. Decay-induced changes therefore
partly cancel the original differences between jets, mainly at small
pair separations. A simple picture comes from a parent paired with
another constituent. Before the decay, the pair contributes to one
shell. The parent's daughters can pair with that constituent at different
separations, spreading the contribution over neighboring shells.
The same fluctuation can then be shared among several shells, with
a smaller amplitude in each. Because variances depend on squared
fluctuation amplitudes, their sum can decrease while the shells
still fluctuate together.
The reduction persists when both stages use the same \(H_0\)
jet-momentum normalization.
A larger particle multiplicity need not imply stronger EEC fluctuations.
Hadronization and decays change how momentum is shared among particles
and how their pair contributions are distributed across angles.

\section{Medium modification of angular fluctuations}
\label{sec:medium}

We follow common initial Pythia histories through vacuum and
medium evolution using the color-tracking LBT framework
\cite{Dang:2026LBT}. Pythia generates the hard scattering and
showers to \(Q_M=2\) GeV. LBT propagates the partons through a
static medium at \(T_{\rm med}=0.3\) GeV from \(t=0.6\) to \(4\)
fm/\(c\), with elastic scattering, induced radiation, and recoil
\cite{He:2015LBTElastic,Luo:2024LBTInelastic}. Pythia resumes the
positive-particle shower to \(Q_h=0.5\) GeV, followed by hadronization
and decays. We record the resumed partons \(P\), undecayed
hadrons \(H_0\), and final visible particles \(H\). The vacuum branch
uses the same shower interruption and restart without medium interactions.

The collision energy, jet definition, and EEC grid follow
Sec.~\ref{sec:observable}. We first select vacuum \(P\) jets with
\(50\leq\pt<60\) GeV and \(|\eta_J|<0.5\) from the leading two jets.
Using these as references, we require mutually nearest matches within
\(\Delta R_{\rm jet}<0.25\) in both branches at all three stages,
yielding about 5300 matched jets.
Negative LBT records subtract the thermal momentum already carried by
recoil particles. They are hadronized and decayed separately, retaining
their sign in jet recombination. We include them in the EEC pair sum
of Eq.~\eqref{eq:rho} through \(z_i=s_i p_{T,i}/p_{T,J}^{\rm net}\),
where \(s_i=\pm1\) and \(p_{T,J}^{\rm net}\) is the transverse momentum
of the signed four-vector sum before the constituent cut.
Both signs have the same \(p_T>1\) GeV cut. The density can be
negative; \(C\) and \(R\) use positive jet weights.

\begin{figure*}[t]
 \centering\includegraphics[width=.98\textwidth]{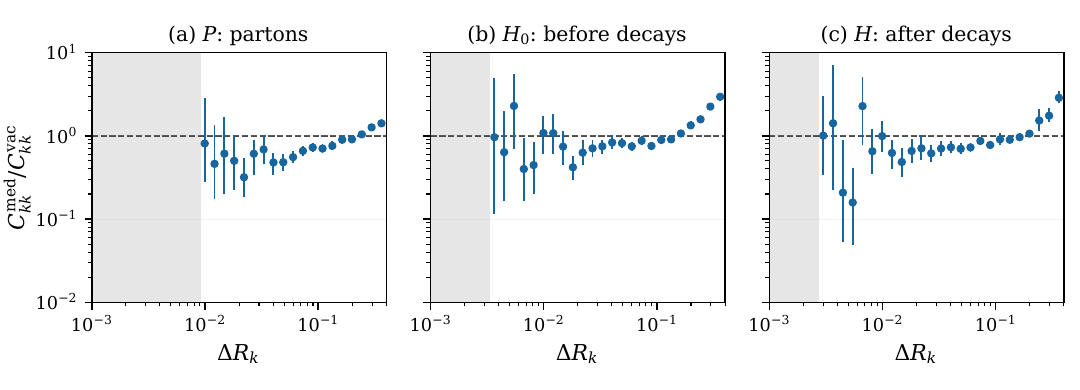}
 \caption{Medium-to-vacuum shell-variance ratios
 \(C^{\rm med}_{kk}/C^{\rm vac}_{kk}\) for the matched jets in
 Fig.~\ref{fig:medium-correlation}. Bars give paired 95\% bootstrap
 intervals; the dashed line marks unity. Gray bins have fewer than
 20 nonzero source events in either branch or a nonpositive vacuum
 variance in any bootstrap replica.}
 \label{fig:medium-variance}
\end{figure*}

The medium weakens the anticorrelation between EEC fluctuations at
smaller and larger pair separations (Fig.~\ref{fig:medium-correlation}). For the final-state
EEC integrals over shells centered in \(A=[0.02,0.1)\) and
\(B=[0.15,0.4)\), including shell widths, the correlation changes
from about \(-0.72\) in vacuum to \(-0.34\) in the medium.
One possible source is the fluctuating momentum transfer to the medium.
The same positive constituent can form pairs with positive partners
at different separations. If that constituent loses momentum,
several pair contributions can decrease together, with the size
of the change varying from jet to jet.
This common response can partly offset the momentum-sharing
competition described in Sec.~\ref{sec:stages}.

The medium affects the size of EEC fluctuations differently at
smaller and larger pair separations. Figure~\ref{fig:medium-variance} shows variance suppression
over much of \(0.01\leq\Delta R<0.1\). For shells centered in this
window \(W\), the final-state medium-to-vacuum ratio of the summed
variance \(S_W=\sum_{k\in W}C_{kk}\) is approximately \(0.66\pm0.07\).
The suppression remains when both branches use the same vacuum-reference
jet momentum for normalization. A possible picture is that scattering
and radiation weaken high-weight close pairs or move them out of a
given shell. If the largest local contributions are reduced, the shell
densities can become less spread out across jets, giving a smaller variance.

At the largest pair separations, the variance ratio instead exceeds
unity. Radiation and recoil can add widely separated pairs, with
different numbers and momenta in different jets. Their contributions
can therefore increase the jet-to-jet spread in the outer shells.
The normalization also matters: at fixed particle momenta, a smaller
jet momentum increases the magnitude of each normalized pair contribution.
Using a common
vacuum-reference momentum reduces the enhancement, showing that the
denominator also affects the outer-shell variance.

\begin{samepage}
Overall, the medium changes not only the size of EEC fluctuations
but also how different angular regions vary together. Both effects remain visible
after hadronization and decays.

\end{samepage}

\section{Conclusions}
\vspace{-0.5\baselineskip}

\begingroup
\widowpenalty=10000
This work extends EEC studies beyond the mean angular profile by
examining jet-by-jet fluctuations and their correlations across angles.
Reconstructed jets in CMS Open Data show large variations in their
EEC profiles, with many empty angular shells and broad, non-Gaussian
shell-density distributions. The covariance and Pearson matrices further
reveal organized angular fluctuations: nearby outer-angle shells tend to vary
together, while separated angular regions are anticorrelated. The
broad variance spectrum shows that these fluctuations involve changes
in the angular shape, not only the overall amplitude.

The Pythia comparison connects these patterns to different stages of
jet evolution. Negative correlations between separated angular regions
are already present at parton level. Momentum sharing provides a simple
picture of this competition: at a fixed momentum budget, changes in the
sharing between branches can increase pair contributions at some
separations while reducing them at others. Hadronization reshapes these
correlations, and decays produce a broader positive band around and
below the mean-EEC turnover. Decays also reduce the summed shell
variance while strengthening nearby-shell correlations. The matched
sample shows that the decay response partly cancels the original
fluctuations, reducing the summed variance by about half. Spreading a
parent's pair contributions across neighboring shells provides a simple
picture of how smaller fluctuations can coexist with stronger correlations.

The matched Pythia--LBT calculation shows that medium effects also
change fluctuation strength and angular correlations in different ways.
In this static-medium calculation, the final state exhibits variance
suppression at smaller pair separations, enhancement in the outer
shells, and weaker anticorrelation between separated angular regions.
The medium-induced changes thus remain visible after hadronization
and decays, in both the size of the fluctuations and their angular
correlation structure.

Together, these results make joint EEC fluctuations a complementary
probe of momentum sharing and pair redistribution in jets. They provide
additional tests of how models distribute particle-pair contributions
across angles and how those contributions vary together from jet to jet.
Jet-by-jet EEC fluctuations thus offer a route to probing the
stochastic evolution of QCD jets.

\endgroup

\vspace{-0.5\baselineskip}
\begin{acknowledgments}
\vspace{-0.5\baselineskip}
The author thanks Dr. Xiaozhou Yu of Fudan University for helpful discussions.

The author used large language model tools, including OpenAI Codex,
for scientific discussions, code development, and manuscript drafting.
\end{acknowledgments}
\vspace{-\baselineskip}
\section*{Data Availability}
\vspace{-0.5\baselineskip}
The collision data are publicly available from the CERN Open Data
Portal \cite{CMS:OpenData2011Jet}. The source-file selection,
derived EEC vectors, numerical results, generator configurations,
and analysis code are available from the author upon reasonable
request.

\onecolumngrid
\clearpage
\twocolumngrid
\appendix
\section{Samples and analysis definitions}
\label{app:samples}

\subsection{CMS Open Data}

The subset of the CMS 2011A Jet dataset
\cite{CMS:OpenData2011Jet} processed with CMSSW 5\_3\_32 yields
54\,906 selected jets from 50\,445 collisions.

We use the certified run and luminosity sections
\cite{CMS:Certified2011}. The first primary vertex must not be
flagged as fake, must have more than four degrees of freedom, and
must lie within
\(|z|<24\) cm and 2 cm of the beam axis in the transverse plane.

We rank jets by uncorrected transverse momentum and retain the
two leading jets before applying the jet-quality and central-acceptance
cuts. The jet-quality cuts require neutral-hadron and
neutral-electromagnetic energy fractions below 0.95, more than
one constituent, nonzero charged-hadron energy and charged
multiplicity, and a charged-electromagnetic fraction below 0.99.

Corrected jet momenta determine the analysis intervals. We use
the \texttt{FT\_53\_LV5\_AN1} AK5PF corrections: L1FastJet,
L2Relative, L3Absolute, and data residual levels \cite{CMS:JEC2011}.
The EEC uses the original PF constituent momenta and the full
uncorrected jet momentum before the constituent cut. All PF
particle types above that cut are included, without constituent-level
pileup subtraction.

Events must pass a version of \texttt{HLT\_DiJetAve30U}.
Each selected jet is weighted by the product of the applicable
L1 and HLT prescales.

\subsection{Generated samples}

The independent three-stage comparison uses Pythia 8.315 with
the Monash 2013 tune
\cite{Bierlich:2022pfr,Bierlich:2022release,Skands:2014pea}.
Each stage contains 600\,000 valid events.
We generate hard QCD \(2\to2\) processes at 7 TeV with
\(\hat p_T>20\) GeV, initial- and final-state radiation, and
multiparton interactions. The tune uses NNPDF2.3 QCD+QED LO
with \(\alpha_s(M_Z)=0.130\).

\begingroup\widowpenalty=10000
The parton sample \(P\) uses final colored partons with
\texttt{HadronLevel:all=off}. For \(H_0\), hadronization is
active and subsequent decays are disabled with
\texttt{HadronLevel:Decay=off}. The \(H\) sample contains final
visible particles after standard decays, including photons and
leptons but excluding neutrinos.
FastJet \cite{Cacciari:2011ma} clusters the particles before the
EEC constituent cut. The separate decay sample described in
Sec.~\ref{sec:observable} contains 600\,000 generated events
and yields 3733 matched jet pairs.\par
\endgroup

The Pythia--LBT study uses 30\,000 common input histories.
Requiring matches across both branches and all three stages
yields 5264 matched jets from 5195 events.

We use 300 bootstrap replicas, with each collision as one
resampling unit. For independent stage samples, each stage's
replica is shared by every ratio containing it. Matched stages
and vacuum--medium branches are resampled together. In the
medium study, events are resampled within each source seed.

\section{Momentum and constituent-threshold checks}
\label{app:checks}

\subsection{Jet momentum}

We test the jet-momentum dependence of the angular correlations
using the other three momentum intervals. Their CMS Pearson
matrices appear in the upper row of Fig.~\ref{fig:cms-checks}.
Together with the 50--60 GeV result in Fig.~\ref{fig:cms-matrix},
they show the separated-angle negative region and nearby
outer-angle positive correlations in all four intervals.

\subsection{Constituent threshold}

For the 50--60 GeV sample, we vary the constituent cut
\(p_{T,i}>t\) over \(t=0,0.3,0.6,1\) GeV while keeping the
selected jets, full-jet normalization, weights, and angular grid
fixed. The lower row of Fig.~\ref{fig:cms-checks} shows the three
thresholds below the nominal 1 GeV cut used in Fig.~\ref{fig:cms-matrix}.
The separated-angle negative region and outer adjacent-bin
positive correlations persist throughout the scan.

We also remove the constituent threshold in the three-stage
Pythia comparison, keeping the selected jets and full-jet
normalization fixed within each stage. Figure~\ref{fig:pythia-zero-threshold}
shows the \(p_{T,i}>0\) counterpart of Fig.~\ref{fig:pythia-matrix}.
The separated-angle negative region and the broader
positive band after decays remain. Some outer-angle central values
of \(C_H-C_{H_0}\) change sign when the cut is removed. Lowering
the cut retains soft constituents, including decay daughters,
and changes the particle pairs entering these covariances.

\onecolumngrid
\clearpage
\noindent\begin{minipage}{\textwidth}
\setlength{\abovecaptionskip}{6pt}
\setlength{\belowcaptionskip}{0pt}
\par\noindent\begin{minipage}[b]{\textwidth}
 \makeatletter\def\@captype{figure}\makeatother
 \centering\includegraphics[width=.82\textwidth]{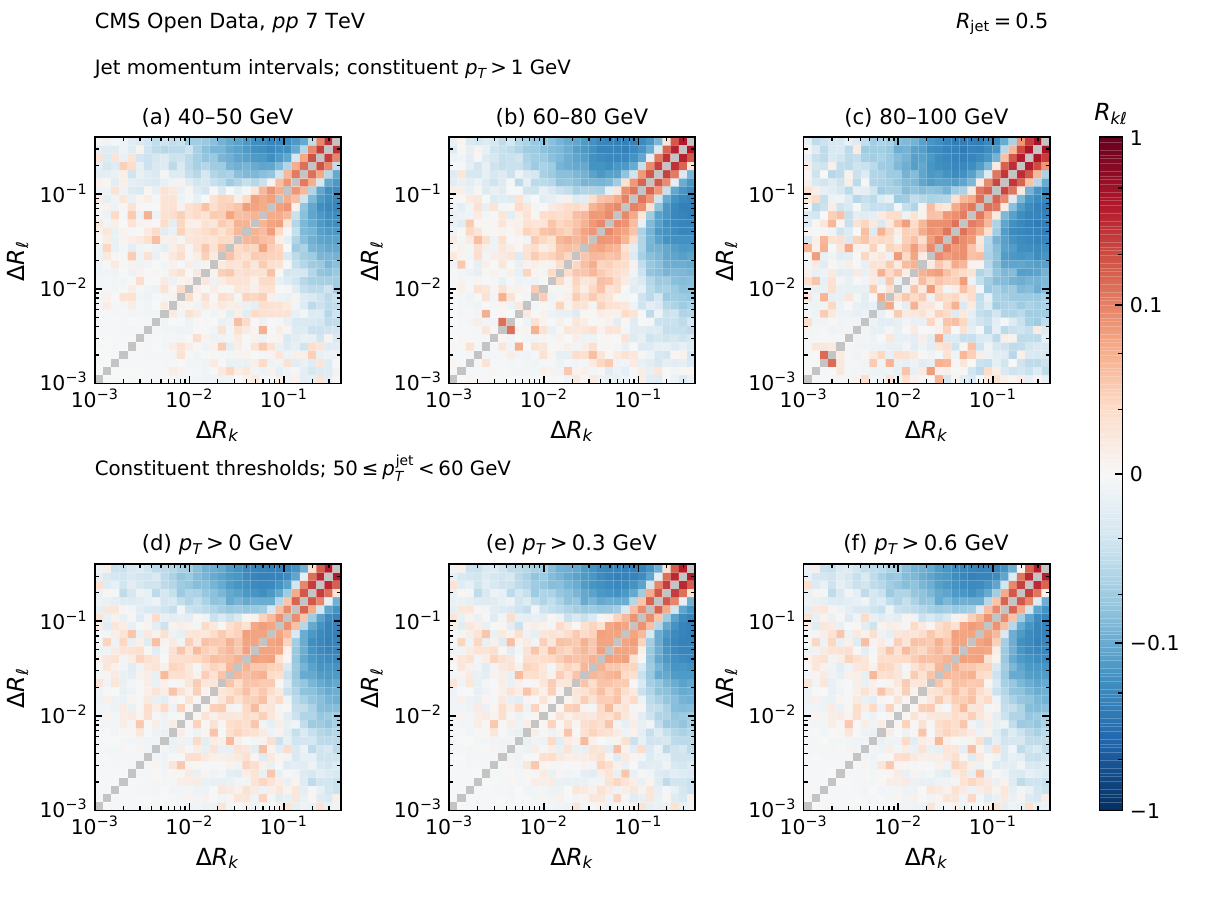}
 \caption{CMS Pearson matrices for jet-momentum (top) and
 constituent-threshold (bottom) checks. Top: 40--50, 60--80,
 and 80--100 GeV jets with \(p_{T,i}>1\) GeV. Bottom: the same
 selected 50--60 GeV jets with thresholds 0, 0.3, and 0.6 GeV.
 Normalization and grid are unchanged; color and masking
 conventions follow Fig.~\ref{fig:cms-matrix}.}
 \label{fig:cms-checks}
\end{minipage}\par

\vspace{8pt}
\noindent\begin{minipage}[b]{\textwidth}
 \makeatletter\def\@captype{figure}\makeatother
 \centering\includegraphics[width=.86\textwidth]{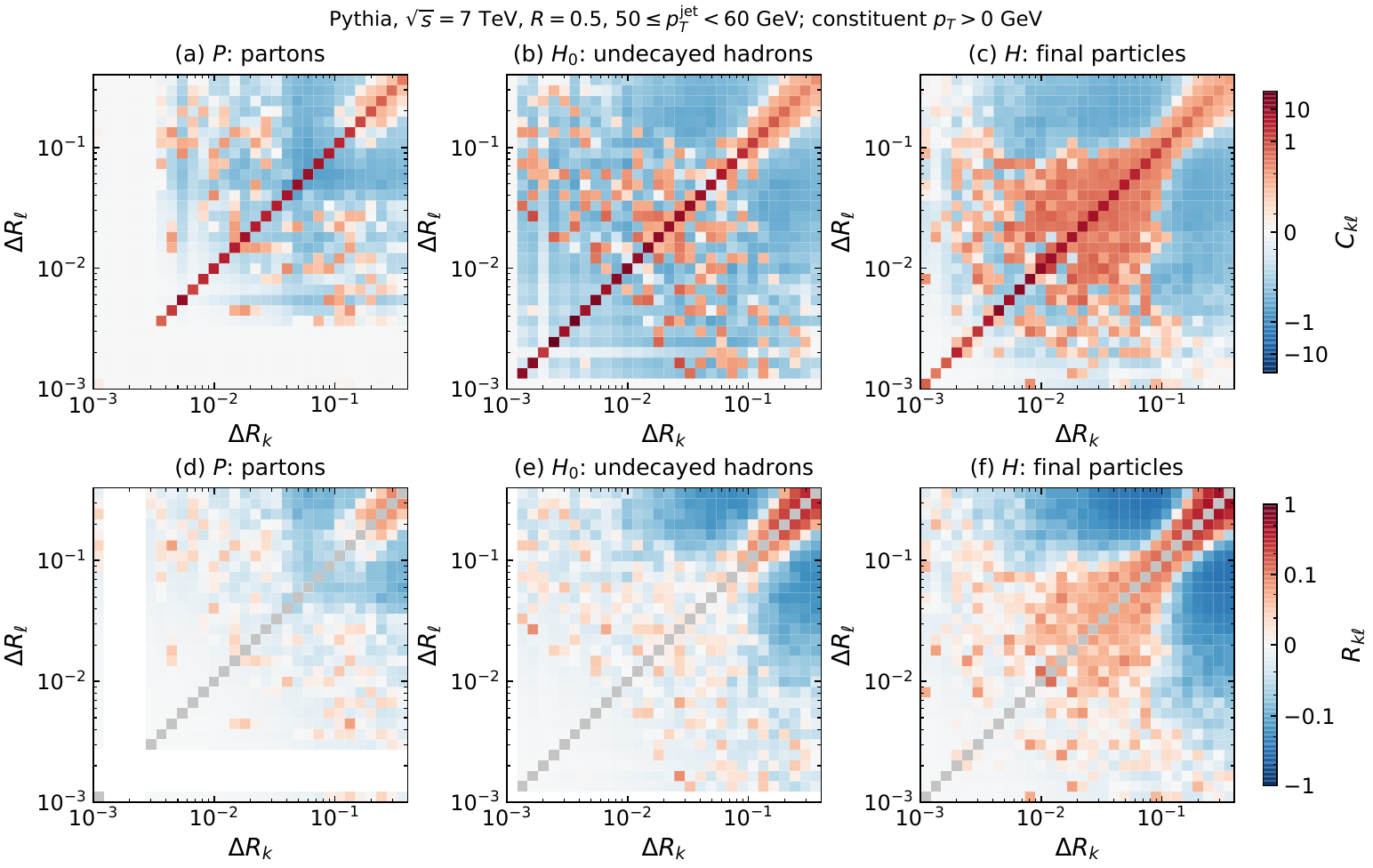}
 \caption{Zero-threshold counterpart of Fig.~\ref{fig:pythia-matrix}
 for 50--60 GeV Pythia jets: \(C\) (top) and \(R\) (bottom),
 with columns \(P,H_0,H\). Only the constituent cut changes to
 \(p_{T,i}>0\); each stage retains its selected jets, full-jet
 normalization, angular grid, and color scales.}
 \label{fig:pythia-zero-threshold}
\end{minipage}\par
\end{minipage}\par
\clearpage
\twocolumngrid

\bibliography{references,cms_references,medium_references}
\end{document}